\documentclass[12pt]{article}

\usepackage{amsmath,amsthm,amssymb}
\usepackage{graphicx}

\newcommand{\1}{\,\mathbf{I}}
\newcommand{\bra}[1]{\left\langle{}#1\right|}
\newcommand{\ddm}{\delta}

\newcommand{\dst}{\mu}
\newcommand{\ee}{\mathbf{e}}
\newcommand{\eff}{E}
\newcommand{\entrp}{S}
\newcommand{\hh}{\mathcal{H}}
\newcommand{\ket}[1]{\left| #1\right\rangle}

\newcommand{\lmu}{\Lambda}
\newcommand{\lth}{n}
\newcommand{\mesb}{\,\rmd\psi}
\newcommand{\pdr}[2]{\frac{\partial\,#1}{\partial\,#2}}
\newcommand{\prxp}[1]{{#1}\otimes{#1}'}
\newcommand{\rrh}{\rho}
\newcommand{\tpar}{B}
\newcommand{\tparsc}{\beta}
\newcommand{\tsth}{H}
\newcommand{\xdm}{\beta}

\DeclareMathOperator{\trc}{Tr} \DeclareMathOperator{\rmd}{d}
\DeclareMathOperator{\rmi}{i} \DeclareMathOperator{\rme}{e}

\newcommand{\bracket}[3]{\bra{#1}#2\ket{#3}}

\newcommand{\tsthp}{\mathbf{\tsth}}

\newcommand{\prjo}[1]{\mathbf{P}_{#1}}

\begin{document}
\title{`Lazy' quantum ensembles}
\author{George Parfionov\thanks{Friedmann Laboratory For Theoretical Physics,
Department of Mathematics, SPb EF University, Griboyedova 30--32,
191023 St.Petersburg, Russia} and Rom\`an Zapatrin\thanks{Department
of Information Science, The State Russian Museum, In\.zenernaya 4,
191186, St.Petersburg, Russia (corresponding author, e-mail
zapatrin@rusmuseum.ru)}}

\maketitle

\begin{abstract}
We compare different strategies aimed to prepare an ensemble with
a given density matrix $\rho$. Preparing the ensemble of
eigenstates of $\rho$ with appropriate probabilities can be
treated as `generous' strategy: it provides maximal accessible
information about the state. Another extremity is the so-called
`Scrooge' ensemble, which is mostly stingy to share the
information. We introduce `lazy' ensembles which require minimal
efforts to prepare the density matrix by selecting pure states
with respect to completely random choice.

We consider two parties, Alice and Bob, playing a kind of game.
Bob wishes to guess which pure state is prepared by Alice. His
null hypothesis, based on the lack of any information about
Alice's intention, is that Alice prepares any pure state with
equal probability. Then, the average quantum state measured by Bob
turns out to be $\rho$, and he has to make a new hypothesis about
Alice's intention solely based on the information that the
observed density matrix is $\rho$. The arising `lazy' ensemble is
shown to be the alternative hypothesis which minimizes the Type I
error.
\end{abstract}


\section*{Introduction}

Consider two parties, Alice and Bob, playing the following game.
Alice prepares a pure quantum state according to certain random
strategy and then sends it to Bob. Initially Bob possesses no
information about Alice's strategy and thus assumes that Alice
performs a completely random choice of pure state, we refer to
this statement as a \emph{null hypothesis}. In this case the
average density matrix received by Bob would be be proportional to
identity.

Measuring the received states, Bob realizes that the average
quantum state emitted by Alice is $\rrh$. However, there are
infinitely many ensembles which average to $\rrh$, and Bob still
can not recover the strategy of Alice. Although Bob now possesses
some information about Alice's intentions: if the received density
matrix $\rrh$ differs from identity, Bob has to make an
alternative hypothesis. To specify such a hypothesis, some extra
principles must be taken into account. These principles should
capture the type of Alice's behavior.

We might assume that the strategy of Alice is to prepare
eigenstates of $\rrh$ with given probabilities, but this is just
an assumption that Alice is `generous' in providing the accessible
information. Or, conversely, Alice might be stingy with the
information and thus chooses pure states according to Scrooge
distribution \cite{josubent}.

In our game, Bob is reluctant to change his opinion and chooses
among Alice's strategies (which average to $\rrh$) the closest to
his null hypothesis. By `closest' we mean minimizing the
Kullback-Leibler \cite{kullback} distance between the
distributions. This distance is the average likelihood ratio and
is associated with the probability of the Type I error\footnote{To
make Type I error means to accept the alternative hypothesis when
the null hypothesis is still valid. An example of Type I error
would be to conclude that the defendant is guilty, when in fact he
or she is innocent.}.

Another way for Bob's reasonings is to assume Alice to be lazy in
efforts to prepare the ensemble. These efforts are quantified in
terms of differential entropy. Remarkably, as we show in Section
\ref{skullback}, both approaches yield the same ensemble
\eqref{egibbs}.

\section{Differential entropy and likelihood ratio}\label{skullback}

First we have to specify a yet vague notion of `preparation
efforts' for an ensemble. Following \cite{shumwestrel} we
formulate it in thermodynamic terms, namely, we quantify these
efforts by the difference between the entropy of uniform
distribution (that is, our null hypothesis) and the entropy of the
ensemble\footnote{We are speaking here of \emph{mixing entropy}
\cite{wehrl} of the ensemble rather than about von Neumann
entropy of its density matrix.} in question. The only obstacle may
occur is to define this entropy, let us dwell on it in more
detail.

The entropy of a finite distribution $\{p_i\}$ is given by Shannon
formula
\[
\entrp(\{p_i\}) = -\sum p_i \ln p_i
\]
This expression diverges for any continuous distribution: we
approximate a continuous distribution $\dst(x)$ by a discrete one
$\{p_i\}$, calculate its Shannon entropy, but it tends to infinity
as we refine the partition. However, we are always interested in
the \emph{difference} between the entropy of the uniform
distribution and the distribution $\dst(x)$ rather then the
entropy itself. At each approximation step we calculate this
difference, and the appropriate limit always exists. To show it
(see, e.g. \cite{handbook} for details), make a partition of the
probability space by $N$ sets $\Delta_i$ having equal uniform
measure. Then the difference $\eff_N$ between the entropies read:
\[
\eff_N \;=\; \ln N \;-\; \left(-\sum p_i \ln p_i\right)
\]
where \(p_i=\int_{\Delta_i}p(x)dx\). The limit expression
$\lim_{N\rightarrow \infty}\eff_{N}$ is the differential entropy

\begin{equation}\label{ediffent}
\entrp(\dst) \;=\; \int \dst(x) \ln \dst(x) \rmd x
\end{equation}

\medskip

\noindent Remarkably, this is equal to Kullback-Leibler distance
\cite{kullback}
\[\entrp(\dst\|\dst_0) \;=\;
\int \dst(x)\ln\frac{\dst(x)}{\dst_0(x)} \rmd x
\] between the distribution $\dst(x)$ and the uniform distribution
$\dst_0(x)$ with constant density, normalize the counting measure
$\rmd x$ on the probability space so that $\dst_0=1$. This
distance is the average likelihood ratio, on which the choice of
statistical hypothesis is based. Then, in order to minimize the
Type~I error we have to choose a hypothesis with the smallest
average likelihood ratio.

\section{`Lazy' ensembles}

The main problem reduces to the following. For given density
matrix $\rrh$ find a continuous ensemble $\mu$ having minimal
differential entropy \eqref{ediffent}:

\begin{equation}\label{egenprob}
\entrp(\dst) \;\to\; \min,\qquad \int \prjo{\psi}\, \dst(\psi)
\mesb \;=\; \rrh
\end{equation}

\noindent  where $\mesb$ is the unitary invariant measure on pure
states normalized to integrate to unity. When there is no
constraints in \eqref{egenprob}, the answer is
straightforward---the minimum (equal to zero) is attained on
uniform distribution. To solve the problem with constraints, we
use the Lagrange multiples method. The appropriate Lagrange
function reads:

\[
\mathcal{L}(\dst) \;=\; \entrp(\dst)\;-\; \trc\,\lmu\left(\int
\prjo{\psi}\, \dst(\psi) \mesb \;-\; \rrh\right)
\]

\noindent where the Lagrange multiple $\lmu$ is a matrix since the
constraints in \eqref{egenprob} are of matrix character.
Substituting the expression \eqref{ediffent} for $\entrp(\dst)$
and making the derivative of $\mathcal{L}$ over $\dst$ zero, we
get

\begin{equation}\label{egibbs}
\dst(\psi) \;=\;
\frac{\rme^{-\,\trc\tpar\prjo{\psi}}}{Z\left(\tpar\right)}
\end{equation}

\noindent where $\tpar$ is the optimal value of the Lagrange
multiple $\lmu$ which we derive from the constraint
\eqref{egenprob} and the normalizing multiple

\begin{equation}\label{epartfun}
Z(\tpar) \;=\; \int \rme^{-\,\trc\tpar\prjo{\psi}} \mesb
\end{equation}

\noindent is the partition function for \eqref{egibbs}.
Substituting he resulting density \eqref{egibbs} to the expression
\eqref{ediffent} for differential entropy we get

\begin{equation}\label{eresentr}
\entrp
\;=\;
\trc \tpar\rrh
\;-\;
\ln Z
\end{equation}

\section{Explicit expressions}\label{sexplexpr}

First evaluate the partition function \eqref{epartfun} in the
eigenbasis of $\tpar$. This integral is a special case of the
calculations carried out in \cite{krwjones}, according to which
$Z(\tpar)$ reads:

\begin{equation}\label{ezb}
Z(\tpar) \;=\; -(\lth-1)!\;\sum_{k=1}^{\lth}
\frac{\rme^{-b_k}}{\prod_{j\neq k}\limits (b_k-b_j)}
\end{equation}

\noindent where $b_k$ are the eigenvalues of $B$. If two or more
of them are equal, the appropriate expression is obtained as a
limit starting with unequal eigenvalues. To write down the
expression for the eigenvalues $\lambda_s$ of the density matrix
$\rrh$ via $\tpar$ we could evaluate the integrals

\[
\lambda_{s} \;=\; \bracket{\ee_s}{\int \prjo{\psi}\, \dst(\psi)
\mesb}{\ee_s}
\]

\noindent in the eigenbasis of $\rrh$. Although, like in
thermodynamics, we have

\begin{equation}\label{edzdb}
\rrh \;=\; \pdr{\ln Z}{B}
\end{equation}

\noindent which gives the explicit expression for the eigenvalues
of the density matrix $\rrh$:

\begin{equation}\label{egrand}
\lambda_{s} \;=\;
-\frac{\frac{\rme^{-b_s}}{\prod_{\stackrel{j=1}{j\neq{}s}}^{\lth}\limits(b_s-b_j)}
\;+\; \sum_{\stackrel{k=1}{k\neq{}s}}^{\lth}\limits
\frac{1}{b_s-b_k} \cdot \left(
\frac{\rme^{-b_s}}{\prod_{\stackrel{j=1}{j\neq{}s}}^{\lth}\limits(b_s-b_j)}
+
\frac{\rme^{-b_k}}{\prod_{\stackrel{j=1}{j\neq{}k}}^{\lth}\limits(b_k-b_j)}
\right)}{\sum_{k=1}^{\lth}\limits
\frac{\rme^{-b_k}}{\prod_{\stackrel{j=1}{j\neq{}k}}^{\lth}\limits(b_k-b_j)}
}
\end{equation}

\noindent from which we see that the resulting density matrix
$\rrh$ remains unchanged when we add a constant to all $b_k$-s.
That means that the matrix `temperature' parameter $\tpar$ for the
lazy ensemble is defined up to an additive constant (in contrast
with classical thermodynamics).

\medskip

\noindent Like in \cite{josubent}, the expression \eqref{ezb} for
the partition function can be given the following integral form

\begin{equation}\label{ezcauchy}
Z(B) \;=\; -\,\frac{({\lth}-1)!}{2\pi\,\rmi} \oint
\frac{\rme^{-z}\,\rmd z}{\mbox{det}(B-z\1)}
\end{equation}

\noindent where the contour encloses all eigenvalues of $\tpar$.

\bigskip

So, given a lazy ensemble \eqref{egibbs} with the parameter
$\tpar$, we have written down the expression \eqref{edzdb} for its
average density matrix. This expression is well-defined for any
matrix $\tpar$. The existence problem remains: given a density
matrix $\rrh$, is there a lazy ensemble with appropriate parameter
$\tpar$ which averages to $\rrh$? Similar question---the existence
of temperature function---arises in thermodynamics. The idea to
solve it is the following \cite{handbook}: we consider the
$\lth$-dimensional CDF (cumulative density function) of the
measure $\dst$ and study the asymptotics of its Laplace transform.
As a result, it can be shown that $\tpar$ exists for any
\emph{full-range} density matrix $\rrh$.

\section{Special case: spin-1/2 particle}

In this case the state space has dimension 2. Write down the
parameter $\tpar$ in the eigenbasis of the density matrix $\rrh$
in a suitable form:

\begin{equation}\label{enewbx}
B\;=\; b\;\cdot\1+\; \left(
\begin{array}{cc}
  -\xdm & 0 \cr
  0 & +\xdm
\end{array}
\right)
\end{equation}

\noindent Then the expression \eqref{ezb} for partition
function reads:

\begin{equation}\label{ez2b}
Z\;=\; \rme^{-b} \cdot \frac{\rme^{\xdm}-\rme^{-x}}{2\xdm} \;=\;
\rme^{-b}
\cdot
\frac{\sinh\,\xdm}{\xdm}
\end{equation}

\medskip

\noindent Calculating the partial derivatives according to \eqref{egrand},
we get the following expressions for the coefficients
$\lambda_{1,2}$ of the density matrix

\begin{equation}\label{ep12}
\lambda_{1,2} \;=\; \frac{1}{2} \;\pm\; \frac{1}{2} \left(
\coth\,\xdm \;-\; \frac{1}{\xdm} \right) \;=\; \frac{1}{2}\pm\ddm
\end{equation}

\noindent where

\begin{equation}\label{edefg}
\ddm \;=\; \frac{1}{2} \left( \coth\,\xdm \;-\; \frac{1}{\xdm}
\right)
\end{equation}

\noindent Denote by $f(\ddm)$ the inverse to $\ddm$. Since the
$\ddm$ is odd and monotone function of $\xdm$, its inverse $f$
exists and bears the same properties. Then the matrix $\tpar$
\eqref{enewbx} is the following function of the density matrix

\[
\tpar \;=\; b\cdot\1\;+\; \left(
\begin{array}{cc}
  f\left(\frac{\lambda_2-\lambda_1}{2}\right) & 0 \cr
  0 & f\left(\frac{\lambda_1-\lambda_2}{2}\right)
\end{array}
\right) \;=\; b\cdot\1\;+\; f\left(\frac{\1}{2}-\rrh\right)
\]

\noindent Since the expression \eqref{edefg} for $\rrh$ is the
independence of the choice of $b$, in two-dimensional case both
matrices $\tpar$ and $\rrh$ are defined by their mean deviation
values $\xdm$ and $\ddm$, respectively. So, the essential
dependence of the matrix `temperature' parameter $\tpar$ from the
density matrix $\rrh$ is completely captured by the function $f$.
Its graph looks as follows.

\begin{center}
\includegraphics[width=6cm]{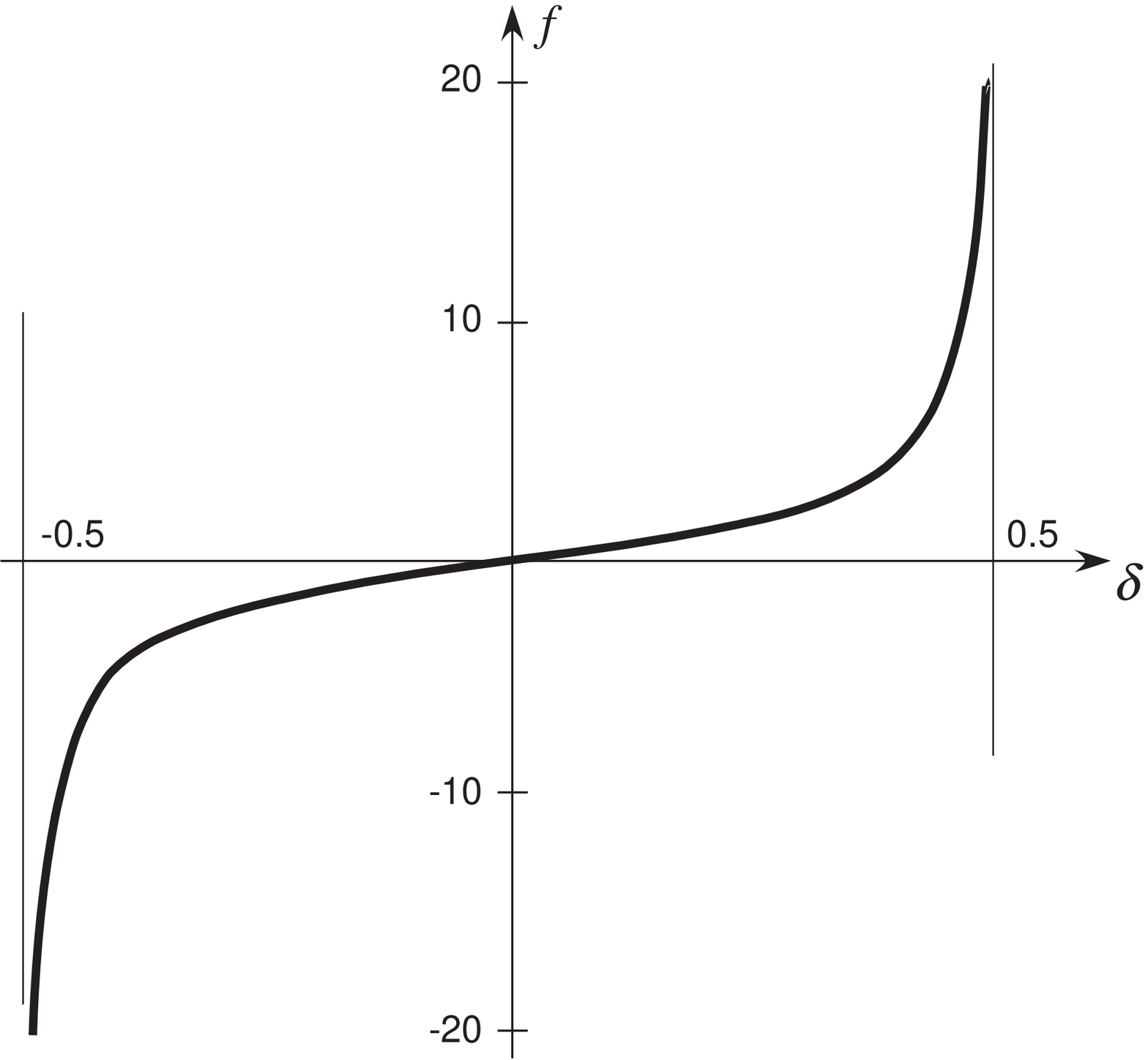}
\end{center}

\section{Lazy ensembles are equilibrium}\label{sequilibr}

Like Gibbs ensembles in thermodynamics, the lazy ensembles are
\emph{equilibrium}, namely, the introduced `temperature'
parameters $\tpar$ possess the equalizing property. To show it,
first introduce the notion of conditional ensemble. In terms of
game played by Alice and Bob this means that Bob measures a fixed
observable $\tsth$ upon the particles emitted by Alice. Again, he
has the uniform distribution as null hypothesis, but the
constraint in \eqref{egenprob} is of scalar rather than of matrix
character. Solving the appropriate variational problem

\[
\entrp(\dst) \;\to\; \min,\qquad \int \prjo{\psi}\, \dst(\psi)
\mesb \;=\; \trc\tsth\rrh
\]

\noindent we obtain

\begin{equation}\label{econdgibbs}
\dst_{\tsth}(\psi) \;=\;
\frac{\rme^{-\,\tparsc\trc\tsth\prjo{\psi}}}{Z_{\tsth}\left(\tparsc\right)}
\end{equation}

\noindent --- this ensemble is \emph{conditional} with respect to
given observable $\tsth$.

\medskip

Consider two quantum systems with state spaces $\hh$ and $\hh'$,
respectively. Let their states initially be $\rrh$ and $\rrh'$.
Then, since we consider a non-interacting coupling of the systems,
the joint density matrix is \(\prxp{\rrh}\) in the tensor product
space \(\prxp{\hh}\). Let us measure the sum of values of the
observables $\tsth$ and $\tsth'$, that is, introduce the
observable \(\tsthp=\tsth\otimes\1'+\1\otimes\tsth'\). The
conditional optimal ensemble of \emph{separable} states with
respect to the observable \(\tsthp\) is the following distribution

\[
\mu_{\tsthp}(\prxp{\psi}) \;=\; \frac{\exp\left[{-\beta_{\tsthp}\,
\trc\tsthp\prjo{\prxp{\psi}}}\right]}{Z_{\tsthp}\left(
\beta_{\tsth}\right)}
\]

\noindent Like in classical thermodynamics, the partition function
of the joint system is the product of subsystems' partition
functions:

\[
Z_{\tsthp}(\tau) \;=\; \int\int
\rme^{-\tau\,
\trc\tsthp\prjo{\prxp{\psi}}} \mesb\mesb' \;=\;\]\[=\; \int\int
\rme^{-\tau\,
\left( \trc\tsth\prjo{\psi}+ \trc\tsth'\prjo{\psi'}\right)}
\mesb\mesb' \;=\; Z_{\tsth}(\tau) \cdot Z_{\tsth'}(\tau)
\]

\noindent therefore the equalizing property holds

\begin{equation}\label{eequrel}
\mbox{If}\quad \beta_{\tsth} \;\le\; \beta_{\tsth'}
\quad\mbox{then}\quad \beta_{\tsth} \;\le\; \beta_{\tsthp} \;\le\;
\beta_{\tsth'}
\end{equation}

\noindent which means that the conditional lazy ensembles are
equilibrium and that $\beta$ plays the r\^ole of temperature
parameter.

\section*{Concluding remarks}

Continuous ensembles of pure states proved their relevance in
various aspects of quantum mechanics. From the theoretical
perspective, they provide the limit cases on which numerical
characteristics of density matrices are attained, for instance,
the minimal value of accessible information about the state is
attained on `Scrooge' ensemble which is a continuous distribution
\cite{josubent}. Furthermore, we claim that they are relevant from
the operationalistic point of view. Even if we are speaking of
preparing discrete ensembles, we must also have in mind that their
are unavoidably smeared by various noises and, strictly speaking,
we have to deal with continuous distributions.

We use the techniques of continuous ensembles to carry out
statistical inference in quantum realm according to the standard
scheme: we have an \emph{a priori} hypothesis (we necessarily need
it, otherwise there is no way to make any inference
\cite{sykora}), then we obtain some information about the system
and have to shift to a new hypothesis.

In our case the null hypothesis is the assumption that any pure
state is emitted with equal probability. Then the information is
obtained that the average density matrix of the state is $\rrh$.
We show how, starting from the `minimal effort' assumption, to
guess the strategy of the preparation of the pure states. As a
result, we obtain so-called `lazy' ensembles.

These ensembles are also proved to provide the minimal deviation
from the null hypothesis. They are described by exponential
distributions \eqref{egibbs} of pure states averaging to a given
density matrix $\rrh$:

\[
\rrh \;=\; \int
\frac{\rme^{-\,\bracket{\psi}{\tpar}{\psi}}}{Z\left(\tpar\right)}
\mesb
\]

\noindent  where the matrix parameter $\tpar$ plays a r\^ole in
some respect similar to temperature, in particular, it is shown to
possess the equalizing property. Although we may not treat it as a
fully-fledged temperature, for instance, in contrast with
classical thermodynamics, it is ambiguously defined up to an
arbitrary additive constant. According to formula
\eqref{eresentr}, we can so choose the additive gauge for $\tpar$
that $\ln Z$ will vanish and the mean value $\trc\tpar\rrh$ will
be equal to the differential entropy of the ensemble, so we may
call this matrix parameter $\tpar$ `differential entropy
observable'.

\paragraph{Acknowledgments.} The authors are grateful to the
participants of A.A.~Friedmann seminar on theoretical physics, in
particular, A.Grib, V.Dorofeev, S.Krasnikov and R.~Saibatalov for
permanent attention to our work and much helpful advice. A support
from RFFI/RBRF --- Russian Basic Research Foundation (grant
04-06-80215a) is appreciated.

\end{document}